# Anisotropic conductance at improper ferroelectric domain walls

D. Meier[1,2,*], J. Seidel[1,3,*], A. Cano[4], K. Delaney[5], Y. Kumagai[6], M. Mostovoy[7], N. A. Spaldin[6], R. Ramesh[1,2,3] & M. Fiebig[6]

[1]Department of Physics, University of California, Berkeley, California 94720, USA

[2]Department of Materials Science and Engineering, University of California, Berkeley, California 94720, USA

[3]Materials Sciences Division, Lawrence Berkeley National Laboratory, California 94720, USA

[4]European Synchrotron Radiation Facility, 6 Rue Jules Horowitz, 38043 Grenoble, France

[5]Materials Research Laboratory, University of California, Santa Barbara, California 93106, USA

[6]Department of Materials, ETH Zürich, Wolfgang-Pauli-Strasse, 8093 Zürich, Switzerland

[7]Zernike Institute for Advanced Materials, University of Groningen, Nijenborgh 4, 9747 AG Groningen, The Netherlands

*These authors contributed equally to this work

**Transition metal oxides hold great potential for the development of new device paradigms because of the field-tunable functionalities driven by their strong electronic correlations, combined with their earth abundance and environmental friendliness. Recently, the interfaces between transition-metal oxides have revealed striking phenomena such as insulator-metal transitions, magnetism, magnetoresistance, and superconductivity[1,2,3,4,5,6,7,8,9]. Such oxide interfaces are usually produced by sophisticated layer-by-layer growth techniques, which can yield high quality, epitaxial interfaces with almost monolayer control of atomic positions. The resulting interfaces, however, are fixed in space by the arrangement of the atoms. Here we demonstrate a route to overcoming this geometric limitation. We show that the electrical conductance at the interfacial ferroelectric domain walls in hexagonal $ErMnO_3$ is a continuous function of the domain wall orientation, with a range of an order of magnitude. We explain the observed behaviour using first-principles density functional and**

**phenomenological theories, and relate it to the unexpected stability of head-to-head and tail-to-tail domain walls in ErMnO$_3$ and related hexagonal manganites[10]. Since the domain wall orientation in ferroelectrics is tunable using modest external electric fields, our finding opens a degree of freedom that is not accessible to spatially fixed interfaces.**

There have been a number of recent reports of unexpected electrical properties at ferroelectric domain walls, particularly in multiferroic materials with their simultaneous ferroelectric and magnetic order[11,12]. For example, high local electrical conductance was measured at ferroelectric domain walls in BiFeO$_3$ whereas the 180° domain walls in hexagonal YMnO$_3$ were found to have higher resistivity than the bulk material[10,13,14]. It is known that such multiferroics often have unconventional mechanisms driving the formation of their ferroelectric domains and domain walls, which are distinct from those of textbook ferroelectrics such as BaTiO$_3$[15]. Multiferroics are therefore a likely source for novel domain and domain-wall properties. In the specific case of the hexagonal manganite multiferroics studied here, the primary symmetry-lowering order parameter is a unit-cell-tripling distortive mode and the subsequent geometrically-driven ferroelectricity is improper[16,17]. The orientation of the resulting spontaneous polarisation is set by the tripling mode, which does not itself introduce a ferroelectric polarisation. It has been shown previously that this is responsible for the unusual distribution of ferroelectric domains[10,18]; here we show that it also has a remarkable impact on the electronic properties of the domain walls.

We choose ErMnO$_3$ as our model hexagonal manganite, although we expect similar behaviour in the other members of the series. Ferroelectricity in ErMnO$_3$ emerges at $T_C \cong 833$ K[19]. Six trimerization-polarisation domains are formed with a spontaneous polarisation $\pm P_z$ pointing along the hexagonal $z$-axis as reported previously for YMnO$_3$[10,18]. In spite of the uniaxial 6mm point symmetry the domains possess a nearly isotropic three-dimensional distribution with a domain size in the order of 1 μm. For our studies platelets with a thickness of about 200 μm and a lateral extension of typically 1×3 mm$^2$ were cut and chemically-mechanically polished with a silica slurry. The platelets were oriented perpendicular to the *z*- or *x*-axis so that they possessed an out-of-plane and an in-plane polarisation, respectively[18]. Samples were investigated by conductive atomic force microscopy (c-AFM) and piezo-response force microscopy (PFM) at ambient conditions.

Figure 1a, b, and c show c-AFM measurements at different bias voltages. They were obtained on the z-oriented sample in a region remote from any domain wall intersection. The image series reveals that the obtainable c-AFM contrast is bias dependent. At −1 V opposite domains with +$P_z$ and −$P_z$ are clearly distinguishable and exhibit different local conductance[20] (see Fig. 1a). The c-AFM brightness changes across $w_{struct}$ ~ 30 nm at each of the two corresponding domain walls. It is important to note that such a change in contrast constitutes an upper limit for the wall width, i.e. the length scale on which the polarisation reverses. At a bias voltage of about −1.6 V regions separating opposite domains become visible. These regions encase the domain walls centrically and possess a conductance lower than that within the domains. With further increase of the bias voltage the width of these low-conductance regions increases to $w_{dress}$ ~ 140 nm at −3.5 V. This strikingly exceeds calculated values of the domain wall width generally reported for ferroelectrics[21,22,23].

Thus, we suggest that two characteristic length scales should be distinguished[24]. We have (i) the usual "structural wall" defined by the reversal of the spontaneous polarisation and associated with the characteristic displacements of ions. Our measurements indicate that the width $w_{struct}$ < 30 nm. In addition we have (ii) an "electrically dressed wall" defined by the region in which the conductance deviates from that within the bulk domains. The width of this region, $w_{dress}$ ~ 10$^2$ nm, is bias-dependent. Our calculations (an analytical solution of the

Poisson equation combined with a numerical solution by the finite-element method) indicate that it is determined by accumulation/depletion effects of itinerant charge carriers and the resulting nontrivial landscape of conductance in the vicinity of the domain wall (see Supplementary Information). In Fig. 1e we compare the current-voltage characteristics of opposite polarisation domains and of the electrically dressed walls separating them. We find that the conductance of the domains can be described by interface-limited Schottky conduction[25] (see fits in Fig. 1e), whereas the dressed walls exhibit a more complex current-voltage characteristic with a plateau from −1.6 to −2.4 V. As consequence, a crossover near −1.9 V occurs (Fig. 1e inset) where the conductance of the dressed walls falls below that of the bulk domains; this is the bias voltage where they become observable.

In order to understand the phenomenon of electrical dressing of domain walls, we repeated the bias-dependent conductance measurements in the highly anisotropic $yz$-plane, i.e., in the plane of the spontaneous polarisation. We find strikingly different conductance behaviour. Figure 2a shows the PFM image of a region within the $yz$-plane of our $ErMnO_3$ sample, with the black arrows indicating the orientation of the spontaneous polarisation within the plane. The characteristic intersection of six domains described previously[10,18] is clearly visible. The inset to Fig. 2a shows the c-AFM data for the same region. It reveals that the domain walls meeting at the intersection have different conductance properties being reflected by the different brightness of the c-AFM signal. Figure 2b shows a larger surface area including two intersections. We see that the c-AFM signal along the walls changes smoothly between dark and bright across length scales of a few micrometers. The brightest signal (high conductance) is found at walls where the polarisation meets tail-to-tail. In contrast, the darkest signal (low conductance) occurs at walls in head-to-head polarisation configuration. In Fig. 2c we compare the bias-dependent conductance at points on the six domain walls leading to the upper intersection in Fig. 2b with the conductance of a bulk domain. The two points in tail-to-tail regions are more conductive (red curves), and the four points in head-to-head regions are less conductive (purple and green curves) than the bulk. In Fig. 3a we present the angular dependence of the normalized local current measured along a "domain wall loop" such as shown in Fig. 2b. Here, $\alpha$ represents the angle between the local wall normal and the direction of the ferroelectric polarisation $P_z$ so that $\alpha = 0°$ corresponds to a head-to-head wall and $\alpha = 180°$ to a tail-to-tail wall (see Supplementary Information for details).

We find a distinctly non-linear relationship between the orientation and the conductance of the domain wall. The difference in conductance at head-to-head and tail-to-tail domain walls can be understood simplistically using the schematics in Fig. 4. At tail-to-tail domain walls (Fig. 4a), the adjacent bound negative charge layers cause an energetically costly divergence in electrostatic potential that must be ameliorated by the accumulation of positive charge. Since ErMnO$_3$ is a p-type semiconductor[10,20,26] this can be achieved using the readily available mobile hole carriers. In contrast, at head-to-head walls (Fig. 4c) the bound positive charge layers require negative charge for screening. Since free negative charges (electrons) are scarce in a p-type semiconductor, screening can only be achieved by bound charges such as cation vacancies, which do not contribute to conductance. At parallel domain walls there is no polar discontinuity and the background conductance of the system is expected (Fig. 4b).

The orientation-dependent spatial distribution of conductivity around charged domain walls in semiconductors was recently predicted in an equilibrium model by Eliseev et al. [27]. Here we adapt their model in order to verify consistency with our results. We find, however, that in order to reproduce our data we need to go beyond the calculation of the equilibrium conductivity and determine the conductance, based on the actual, non-equilibrium flow of current from the c-AFM tip through the sample, both within and outside the region defining the domain wall.

The nominal conductance at the domain wall is expected to be proportional to its density of holes $p$: $\sigma_{DW} \sim p = p_0\exp(-e\phi_{DW}/k_BT)$ where $p_0$ is the density of acceptors and $\phi_{DW}$ is the electric potential at the wall. As in Ref. [27] this potential can be obtained from Poisson's equation

$$\phi'' \approx -\frac{e}{\varepsilon}(p - p_0) = -\frac{ep_0}{\varepsilon}\left(\exp\left(-\frac{e\phi}{k_BT}\right) - 1\right) \quad (1)$$

which determines the variation of the potential in the direction normal to the wall, and the corresponding boundary conditions: $\phi(\pm\infty) = 0$ and $\phi'(\pm 0) = \mp(P_z/\varepsilon)\cos\alpha$. Note that $2P_0\cos\alpha$ is just the density of bound charges accumulated at the surface of the domain wall[28], and $\phi(0)=\phi_{DW}$. Integrating Eq. 1 one obtains

$$\exp\left(-\frac{e\phi_{DW}}{k_BT}\right) + \frac{e\phi_{DW}}{k_BT} - 1 = \frac{1}{2}\Pi^2\cos^2\alpha \quad (2)$$

where $\Pi^2 = P_z^2/(\varepsilon k_BTp_0)$. For realistic values[26] — $P_z = 5$ μC/cm², $\varepsilon = 50\ \varepsilon_0$, and $p_0 = 2 \cdot 10^{19}$ cm$^{-3}$ — we obtain $\Pi \gg 1$ so that Eq. 2 cannot be linearized: The screening of the charged domain walls is a nonlinear phenomenon.

As mentioned above, the equilibrium distribution of $\sigma_{DW}$ does not yet reveal the non-equilibrium current flowing through the sample when we probe the domain structure. The reason is that the current injected by the c-AFM tip will unavoidably spread out beyond the domain wall, thus also probing the conductance of the adjacent region, i.e., beyond $w_{dress}$. In our calculation, the continuity equation $\nabla j = 0$ with the current $j = \sigma\nabla\phi$ is solved for the boundary conditions imposed by the (spatially expanded) tip. The analytical solution (see Supplementary Information) reveals a spreading that is quite obvious for the head-to-head walls (Fig. 3c): They are surrounded by a more conductive environment due the corresponding depletion of charge carriers. However, a strikingly pronounced spread-out of the current is also present for negatively charged walls (Fig. 3d) in spite of their locally enhanced conductance. When we take into account this geometrical effect we can compute the actual orientation-dependent conductance $\sigma_{spread}$ of the domain walls. We obtain

$$\sigma_{spread} \sim \frac{\sigma_{DW}(\alpha)}{1 + \frac{2r}{w(\alpha)} \ln \frac{\sigma_{DW}(\alpha) + \sigma_{bulk}}{2\sigma_{bulk}}} \quad (3)$$

where $\sigma_{bulk}$ is the bulk conductance and $r/w(\alpha)$ is the ratio between the radius of the c-AFM tip and the effective orientation-dependent width $w(\alpha)$ of the domain wall (determined by the accumulation/depletion of holes). This is shown by the red curve in Fig. 3a which qualitatively reproduces the experimental behaviour.

By inserting parameters we can now describe the measured orientation-dependence of $w_{dress}$ as seen in Fig. 3b. A rough estimation indicates that ~ 6 x $10^{13}$ holes per $cm^2$ are required (~0.1 per formula unit) to completely screen the bound charges at tail-to-tail walls when $P_z$ = 5 $\mu C/cm^2$. For the charge carrier density $p_0$ this corresponds to a delocalization of screening charges over ~60 nm which is consistent with the range of values measured for $w_{dress}$.

In order to identify mechanisms that may enhance the orientational variation of the domain wall conductance we performed density-functional calculations of the electronic structure of both head-to-head and tail-to-tail walls of $YMnO_3$, a completely analogous material (see Fig. 4d; details of the calculation and construction of the supercell are provided in the Supplementary Information). Consistent with the electrostatic model just described, the calculations reveal a variation in the electrostatic potential, which reverses at the domain walls with the reversal of polarisation orientation (Supplementary Fig. S3). As shown in Fig. 4d this variation in electrostatic potential shifts the Fermi level at the tail-to-tail walls to lie in the broad O(2p)–Mn(3d) valence band where the carrier effective mass is low and the corresponding hole mobility is high. In contrast, any electrons that accumulate at head-to-head walls can only occupy the narrow $d_{z^2}$ band, where they will have a high effective mass. Thus, in addition to the contribution from the varying carrier concentrations[27] the conductance of the "electrically dressed wall" is further exaggerated by band-structure effects.

In summary, we have demonstrated that the conductance of ferroelectric domain walls in hexagonal ErMnO$_3$ varies continuously with the orientation of the wall. The variation of the conductance by an order of magnitude between head-to-head and tail-to-tail domains walls is the combined consequence of carrier accumulation and band-structure changes at the walls. Both of these effects were derived theoretically, using phenomenological electrostatic and ab-initio density functional theory, respectively.

In conventional ferroelectrics such energetically unfavorable head-to-head and tail-to-tail domain walls are usually avoided. However, their presence is enforced in ErMnO$_3$ because of the protected topology of intersecting domain states in hexagonal manganites[10,18]. Since ferroelectric domain walls, in particular those with an intrinsically isotropic orientation as in the present case, can be routinely modified using electric fields, our results suggest a new degree of flexibility for domain boundary engineering[29] and oxide-based devices, in which interfaces can be dynamically modified even after assembly into a device architecture.

# Figure Legends

**Figure 1 | Bias-dependent domain and domain wall conductance. a**, **b**, **c**, Local c-AFM image series with different bias voltages across a $+P_z$ domain with adjacent $-P_z$ domains acquired on a *z*-oriented crystal. Above a threshold of −1.6 V electrically dressed domain walls become visible and broaden upon further increase of the bias voltage. **d**, Waterfall plot of the current cross sections showing the evolution of measured local domain and domain wall width as a function of applied bias. **e**, Extracted local conductance. While the domains show Schottky-like conductance (solid lines represent fits), the conductance of the electrically dressed wall exhibits clear deviations leading to a crossover around −2 V as presented in the inset (solid lines are guidelines to the eye).

**Figure 2 | Anisotropic electrical conductance of ferroelectric domain walls. a**, PFM image obtained within the *yz*-plane of an $ErMnO_3$ crystal. The direction of polarisation is indicated by arrows. The inset shows a c-AFM image acquired at the same position. Domain walls appear as lines of different brightness on an otherwise homogeneous background reflecting their different conductance. **b**, c-AFM image of two neighbouring singularities in the *yz*-plane of $ErMnO_3$ with polarisation directions in-plane as indicated. **c**, Extracted local conductance of all domain walls at the positions indicated in the inset. The domain wall conductance can either be higher or lower than the bulk conductance varying over one order of magnitude.

**Figure 3 | Angular dependence of the local domain wall conductance. a,** Angular dependence of the normalized current $(I_{wall}(\alpha)-I_{bulk})/I_{bulk}$ measured along the "domain wall loop" in the center of Fig. 2b (details on the geometry in Supplementary Information). Extrema are found whenever domain walls are oriented perpendicular to the direction of

polarisation which is qualitatively reproduced by the red and blue line plots ( $\prod = 3$ ) showing the calculated angular dependence with and without spreading of the current, respectively (see text for details) **b**, Measured and calculated angular dependence of the domain wall width. A pronounced increase in wall width is observed for the head-to-head domain walls. Note that the red line plots in **a** and **b** are generated on the basis of Eq. 3 using the same set of parameters. **c**, Computed spreading of the current injected by the c-AFM tip at a head-to-head domain wall. Due to the depletion of charge carriers (holes) at the wall the environment is more conductive and the current spreads out. **d**, Computed current distribution at a tail-to-tail wall. Although the conductance is locally enhanced a remarkable leakage out of the domain wall occurs (see Supplementary Information).

**Figure 4 | Electronic structure of charged and uncharged ferroelectric domain walls. a**, **b**, **c** Schematic illustrations displaying tail-to-tail, side-by-side, and head-to-head domain walls, respectively. Here, + and − indicate positive and negative bound charges at the domain wall. A colour scale is used to represent the electrical conductance and the associated hole density (white – high, dark brown – low), see text for details. **d**, Calculated local density of states in the head-to-head (upper right), tail-to-tail (lower right) and domain-centre (middle right) regions of the $YMnO_3$ supercell shown on the left. The supercell was constructed with 4 layers of "up" polarised $YMnO_3$ frozen to the calculated bulk ferroelectric structure, alternating with four layers of "down" $YMnO_3$. The Mn and O ions in the interface layer were placed at their corresponding paraelectric positions, resulting in a mirror plane at the interface. The black lines in the density of states plots indicate the sum of the local density of states, while the blue and red lines show the oxygen and manganese contributions respectively. Note the shift up in energy of the bands from the head-to-head to the tail-to-tail configuration, caused by the gradient in the electrostatic potential (see Supplementary Information).

**Figure 1**

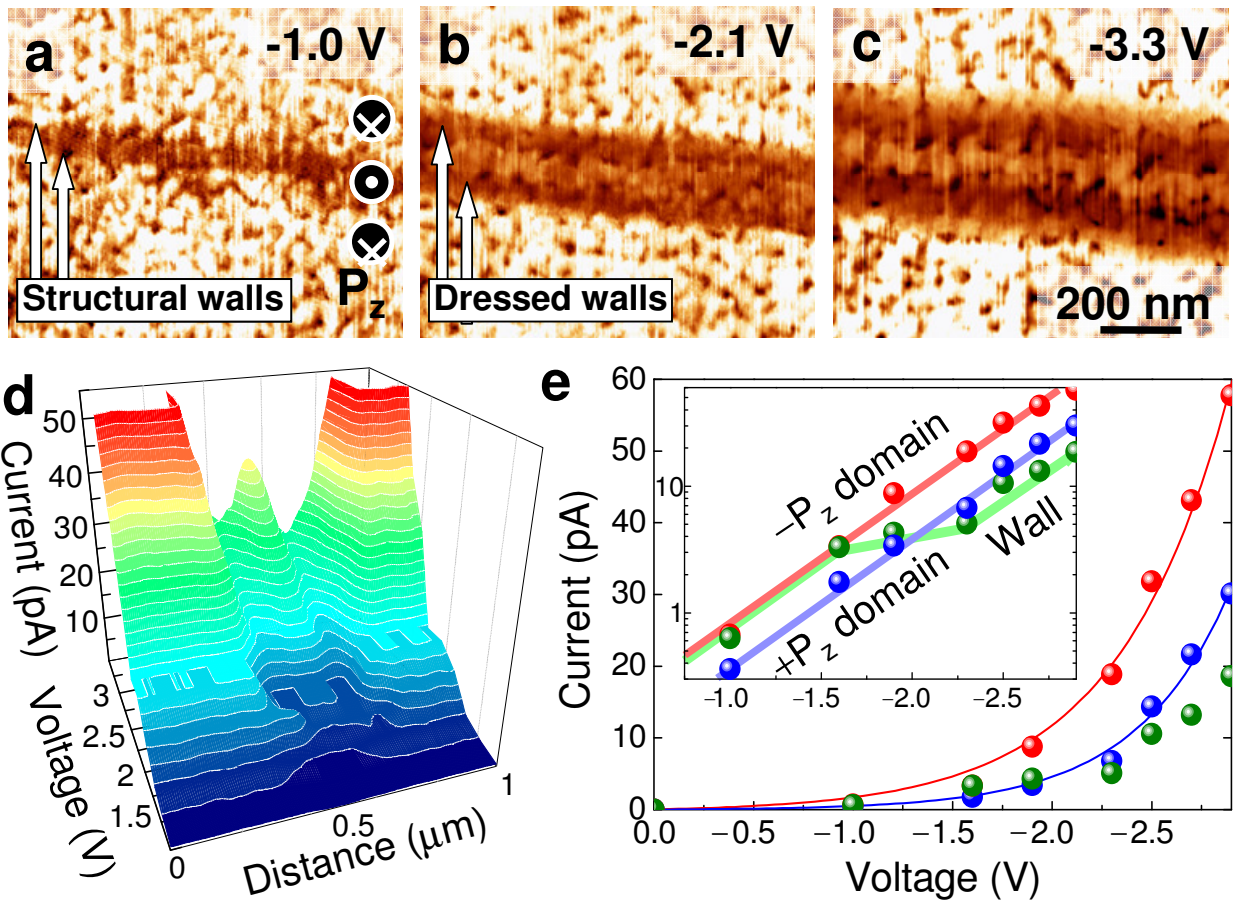

**Figure 2**

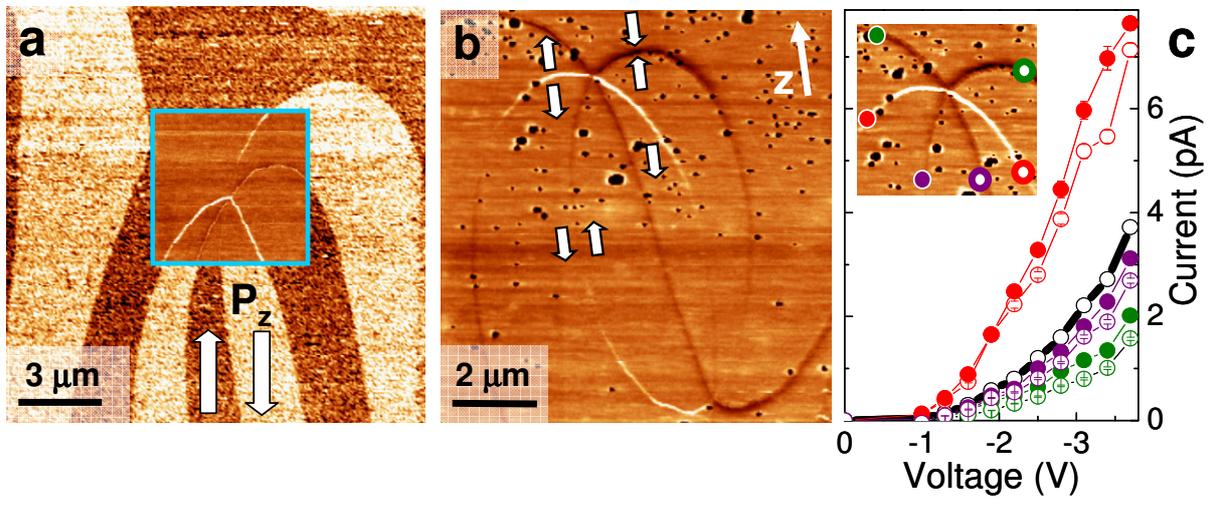

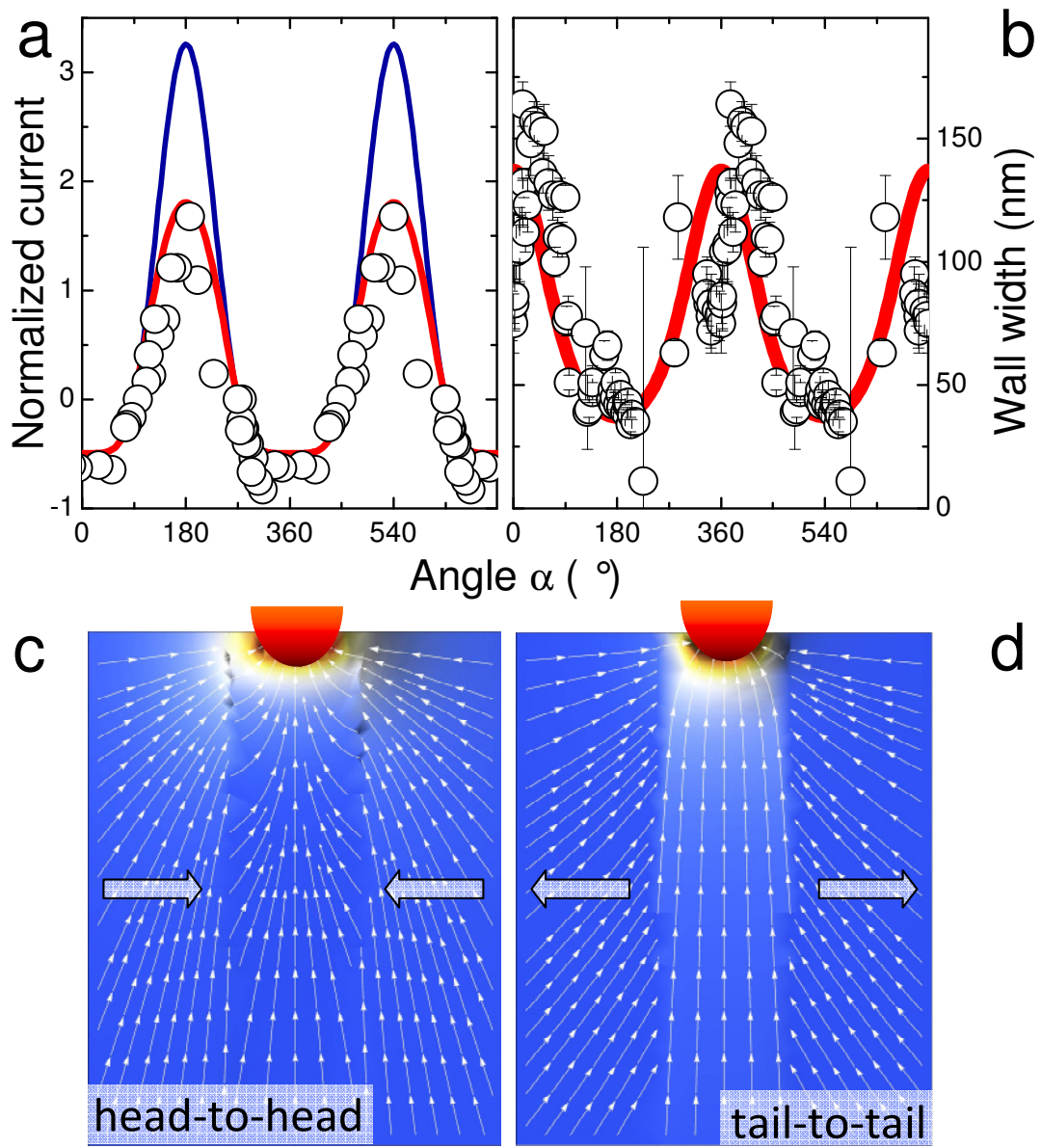

**Figure 4**

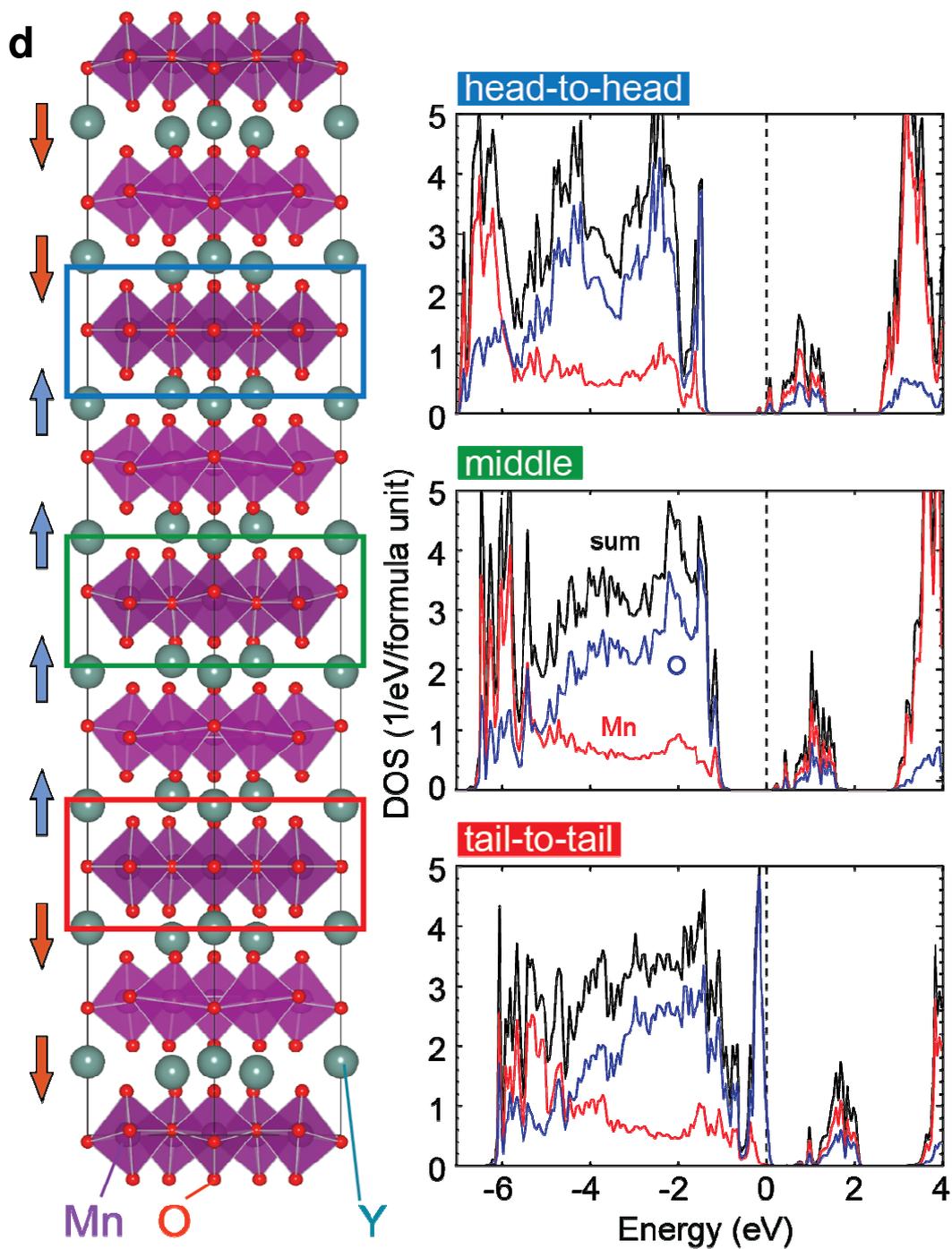

# Supplementary Information

Anisotropic conductance at improper ferroelectric domain walls

D. Meier[1,2,*], J. Seidel[1,3,*], A. Cano[4], K. Delaney[5], Y. Kumagai[6], M. Mostovoy[7], N. A. Spaldin[6], R. Ramesh[1,2,3] & M. Fiebig[8]

[1]Department of Physics, University of California, Berkeley, California 94720, USA
[2]Department of Materials Science and Engineering, University of California, Berkeley, California 94720, USA
[3]Materials Sciences Division, Lawrence Berkeley National Laboratory, California 94720, USA
[4]European Synchrotron Radiation Facility, 6 Rue Jules Horowitz, 38043 Grenoble, France
[5]Materials Research Laboratory, University of California, Santa Barbara, California 93106, USA
[6]Department of Materials, ETH Zürich, Wolfgang-Pauli-Strasse, 8093 Zürich, Switzerland
[7]Zernike Institute for Advanced Materials, University of Groningen, Nijenborgh 4, 9747 AG Groningen, The Netherlands
*These authors contributed equally to this work

## 1. Local *I-V* measurements with negative and positive bias voltage

Figure S1a shows a schematic of the experimental setup. All measurements were performed under ambient conditions using n-doped conductive diamond tips. The network of domains in $ErMnO_3$ prohibits a straight conductance path through the crystal (thickness ~200 microns). Yet, variations in the local conductance are large enough to image the distribution of ferroelectric domains and domain walls at the surface. The electrical circuit from the c-AFM tip to the bottom electrode is in any case closed because of the finite conductance of the sample.

The local *I-V* curves gained on the *x*-oriented sample in the centre of a ferroelectric domain and at domain walls exhibiting head-to-head and tail-to-tail configurations are displayed in Fig. S1b. For negative bias voltages the electronic conductance of tail-to-tail (head-to-head) domain walls appears to be enhanced (suppressed) with respect to the centre of the domain. In contrast, both types of domain wall show equivalent conductance properties for positive bias voltages, at which they are slightly less conductive than the ferroelectric domains. The locally obtained nonlinear *I-V* curves are in good agreement with *I-V* curves extracted from the spatially resolved measurements in Fig. 2c in the main text and indicate the presence of a

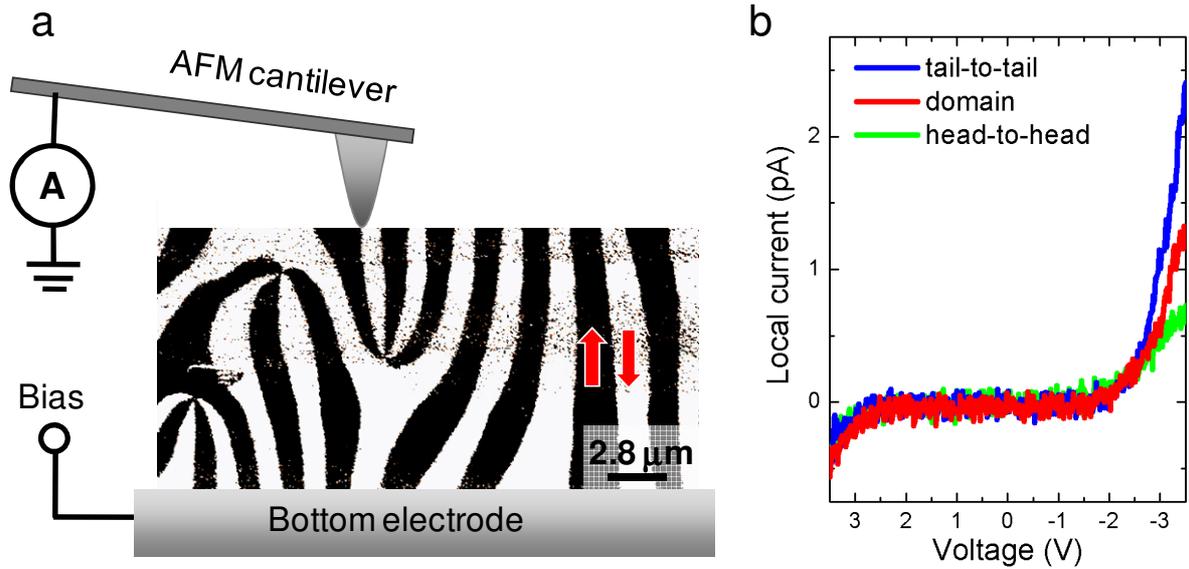

Figure S1. **Schematic setup and local *I-V* curves. a** Setup for c-AFM measurements. **b** Local *I-V* characteristic of head-to-head and tail-to-tail domain walls and of the centre of a domain in ErMnO$_3$.

Schottky-like barrier at the sample-tip interface. In consequence, contrasts are observable in our c-AFM measurements for negative bias directions.

## 2. Angular dependence of the domain wall conductance and width

The angular dependence of the normalized local current and the domain wall width shown in Figs. 2d and 2e of the main text were determined as detailed in the following.

Figures S2a and S2b display examples for a "domain wall loop" covering all the angles between 0° and 360°. The position of the maximum and minimum of the measured currents are marked in Fig. S2a. These extrema correspond to the tail-to-tail and head-to-head polarisation configuration at the wall, respectively.

For the sake of clarity the definition of the wall orientation angle $\alpha$ is presented in Fig. S2c. It is defined as the angle between the z-axis and the normal of the ellipse tangent at the position where the conductance is measured. Note that two cases have to be distinguished: One where the polarisation along the inside contour of the ellipse is parallel to the z-axis and one where it is antiparallel to the z-axis. The two examples in Figs. S2a and S2b both correspond to the parallel case. For an antiparallel alignment a phase shift of 180° has to be considered to sustain a unique relation between the angle $\alpha$ and the type of domain wall (head-to-head wall: $\alpha = 0°$; tail-to-tail wall: $\alpha = 180°$).

This leads to the angular dependence of shown in Fig. S2d. Here, normalized c-AFM currents defined as $(I_{wall}(\alpha)-I_{bulk})/I_{bulk}$ are plotted with $I_{bulk}$ being the bulk background current as averaged from the images. $(I_{wall}(\alpha)-I_{bulk})/I_{bulk}$ is referred to as normalized domain wall current and is used as a measure of the relative conductance with respect to the bulk. Figure S2d shows that similar results are obtained on the loops in Figs. S2a and S2b. The solid lines (red and blue) in Fig. S2d represent the current values observed along closed loops as depicted in Fig. S2a. In contrast, the data points (blue and red circles) are extracted from cross sections perpendicular to the wall. For this purpose a Gaussian profile was fitted to the domain wall cross section so that its maximum can be assigned as the local current and the full width at half maximum as the local domain wall width. Note that independent of the method used for the data analysis qualitatively equivalent results are obtained. A common feature of all data sets is a small shift away from 90° and 270° (side-by-side wall configuration) regarding the zero-crossing of $(I_{wall}(\alpha)-I_{bulk})/I_{bulk}$. This shift can be attributed to local defects, the unknown progression of the domain walls into the sample, and the proximity of a domain wall intersection. The influence of the latter can be seen by comparing the "domain wall loops" in Figs. S2a and S2b.

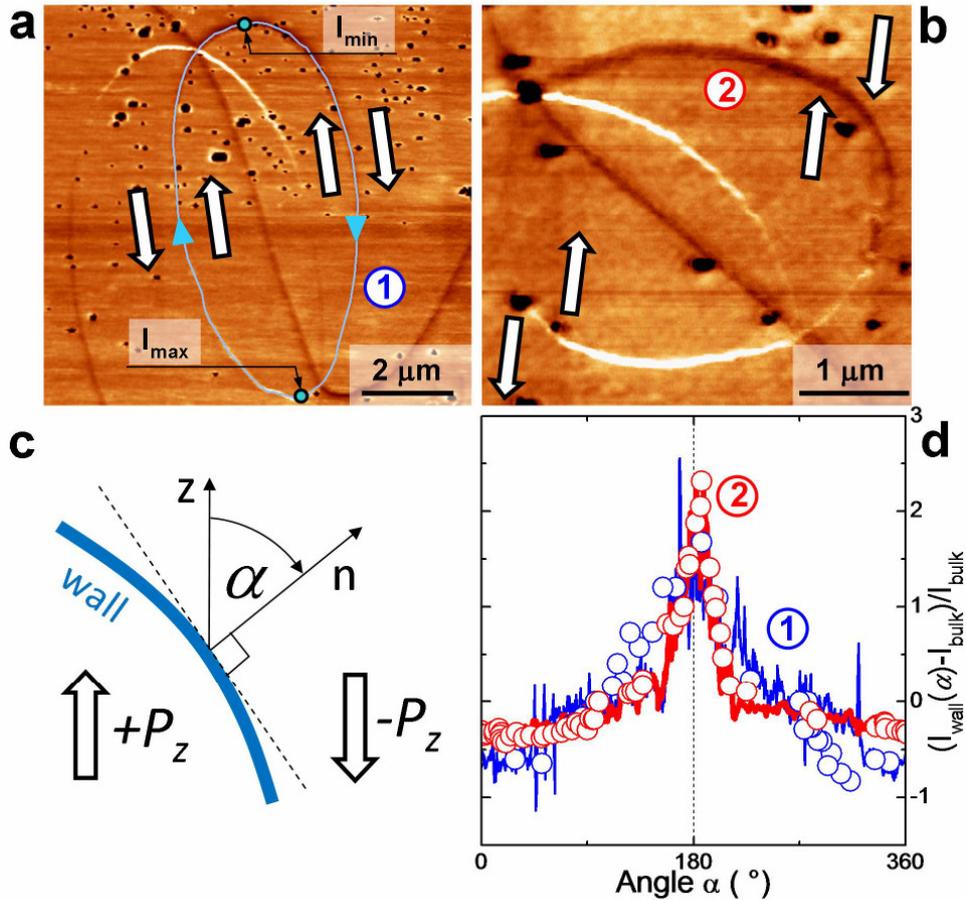

Figure S2. **Angular dependence of domain wall conductance. a** The blue line shows the pathway along which the local conductance has been extracted from the c-AFM image. Maximum and minimum of the observed current are indicated. White arrows denote local polarisation directions of the domains adjacent to the considered "domain wall loop". **b** Another domain wall loop for comparison with panel a. **c** Definition of the domain wall orientation angle. **d** Plots of the normalized angular conductance deduced from Figs. S2a and S2b.

## 3. DFT calculation details

Density functional calculations were performed using the VASP code [G. Kresse and J. Furthmüller, *Phys. Rev. B* **54**, 11169 (1996)] with the GGA+U method in the Dudarev implementation [J. P. Perdew, K. Burke, and M. Ernzerhof, *Phys. Rev. Lett*. **78**, 1396 (1997); S. L. Dudarev et al., Phys. Rev. B 57, 1505 (1998).] and a $U_{eff}$ of 5 eV. PAW potentials with semi-core states in the valence (Y: (4s)2(4p)6(4d)1(5s)2, Mn: (3d)5(4s)2, O: (2s)2(2p)4) were used. Convergence parameters were an energy cutoff 550 eV, SCF energy convergence $10^{-5}$ eV, and a Gamma-centered 6×6×1 k-point grid. We imposed an antiferromagnetic magnetic configuration with alternating ferrimagnetic layers (up-up-down, down-down-up, etc.). With these parameters the band gap of ideal bulk $YMnO_3$ at the experimental lattice parameters and ionic positions is 1.35 eV.

To model the head-to-head and tail-to-tail domain walls, we constructed a supercell containing four layers of up-polarised $YMnO_3$ at the experimental positions of the ferroelectric structure [Gibbs *et al.*, *Phys. Rev. B* **83**, 094111 (2011)] alternating with four layers of down-polarised along the z direction. The positions of the oxygen and manganese ions in the domain walls were set to the paraelectric values and mirror symmetry was imposed. The resulting unit cell, with 120 atoms, contained one head-to-head and one tail-to-tail domain wall. The ions were fixed at these positions and not allowed to relax. With this configuration the variation in electrostatic potential (Figure S3) across the cell was found to be ~1 eV; in supercells with fewer polar layers and/or reduced magnitudes of the polarisation we obtained a correspondingly smaller variation in the electrostatic potential. The shifts in the band edges shown in Fig. 3d of the main paper are in concordance with the calculated variation in electrostatic potential.

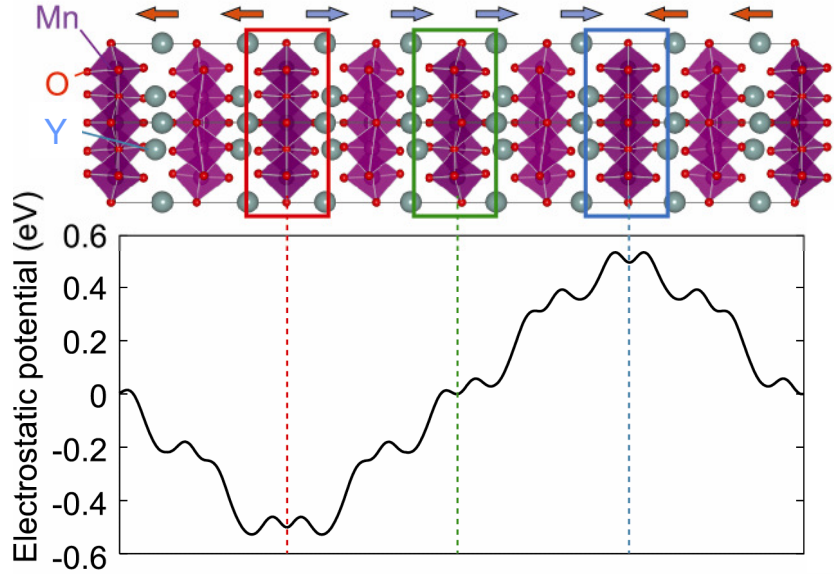

Figure S3. **Calculated electrostatic potential at a tail-to-tail (red) and head-to-head (blue) domain walls.** Note the change in slope of the potential at the walls and its change in sign in the mid-domain region (green).

To test the robustness of our conclusions, we repeated our calculations using the ABINIT code [X. Gonze et al., *Comput. Mater. Sci.* **25**, 478 (2002)] and a smaller (1/3 the size) in-plane unit cell which neglected the tiltings of the $MnO_5$ polyhedra. While this structure and code resulted in a smaller band gap, our results yielded qualitatively the same physical picture.

## 4. Charge redistributions and surface conductance

The distribution of the current injected by the c-AFM tip into the sample can be obtained from the equation of continuity $\nabla \cdot j = 0$ under the corresponding boundary conditions imposed by the tip. We follow a perturbative approach, in which the current density is given by $j = \sigma \nabla \phi$, where $\sigma$ is the sample conductivity in absence of bias (determined by the equilibrium distribution of charge carriers) and $\phi$ is the electric potential due to the tip.

In our measurements, we are dealing with interfaces in which there is a redistribution of mobile charge carriers due to the bound charges that accompany the abrupt (longitudinal) variations of the electric polarization from $+P_z$ to $-P_z$. When the c-AFM scans are performed on a *z*-oriented crystal, this redistribution takes place across the entire *xy*-surface of the sample. On the other hand, when the *yz*-plane is scanned, such redistribution occurs in the vicinity of the domain walls only.

For scans of *z*-oriented crystals, for example, the conductance of the sample can be taken as

$$\sigma = \begin{cases} \sigma_{surf} & -d < z < 0 \\ \sigma_{bulk} & z < -d \end{cases} \quad (S1)$$

where $\sigma_{bulk}$ represents the nominal conductance of the system and $\sigma_{surf}$ the average conductance at the surface obtained from the corresponding accumulation/depletion of charge carriers over the distance $d$. Both $\sigma_{surf}$ and $d$ obviously depend on the bound charges that appear at the surface (that is, on the spontaneous polarization $P_z$). In this model, the exact distribution of current injected by the c-AFM tip can be obtained analytically from the potential:

$$\phi_0(x,y,z) = \begin{cases} \dfrac{I}{2\pi\sigma_{surf}} \sum_{n=-\infty}^{\infty} \dfrac{f^{|n|}}{\sqrt{\rho^2 + (z-2nd)^2}} & -d < z < 0 \\ \dfrac{I}{2\pi\sigma_{bulk}} \sum_{n=0}^{\infty} \dfrac{gf^n}{\sqrt{\rho^2 + (z-2nd)^2}} & z < -d \end{cases} \quad (S2)$$

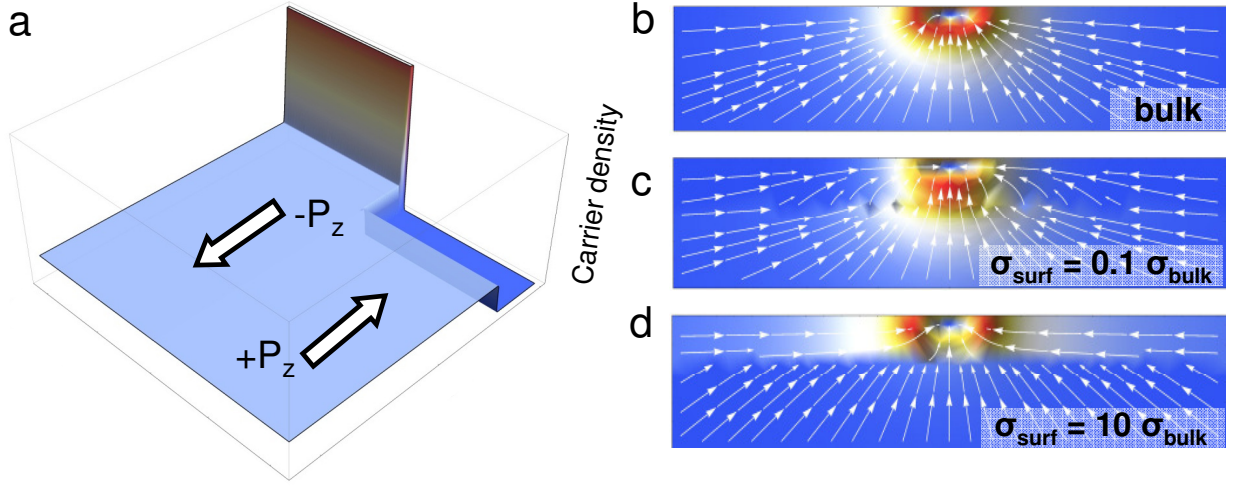

Figure S4. **Calculated surface charge carrier density and related current distribution. a** Landscape of conductance on the surface of a *z*-oriented sample. Mobile charge carriers (holes) accumulate on the $-P_z$ domains to screen the bound charges at the surface leading to enhanced conductance. In contrast, a depletion occurs at the surface of the $+P_z$ domains so that the conductance is reduced compared to the bulk. **b, c, d** Distribution of the current injected by the c-AFM tip for the bulk, $+P_z$ domains, and $-P_z$ domains, respectively (see text for details).

Where $\rho^2 = x^2 + y^2$, $f = (\sigma_{surf} - \sigma_{bulk})/(\sigma_{surf} + \sigma_{bulk})$, and $g = 2\sigma_{surf}/(\sigma_{surf} + \sigma_{bulk})$. This potential corresponds to the injection of current by a point source placed at the surface of the sample (at the position $x = y = 0$). The c-AFM tip can be seen as a distribution of such point sources over the surface of the sample and, accordingly, the resulting potential and distribution of current can be obtained as the corresponding sum (strictly speaking an integral) over these sources.

Our results are summarized in Fig. S4 and lead to the following explanation for the domain wall width observed in scans on the *z*-oriented sample as shown in Figs. 1a, b, and c.

Since the system is a p-type semiconductor, itinerant charge carriers (holes) are expected to accumulate at the surface of the $-P_z$ domains while they will deplete at the surface of the $+P_z$ domains. Thus, the carrier density changes from accumulation to depletion across the domain wall giving rise to the nontrivial landscape of conductance shown in Fig. S4a. Such a landscape, however, does not suffice to explain the width of the electrically dressed walls since the change from accumulation to depletion takes place within a distance in the order of the Debye length which can be estimated as 2 nm in our system.

The key ingredient is the distribution of current injected by the c-AFM tip. The distribution is such that, when the conductance at the surface is lower than underneath in the bulk, the region that is actually probed by the c-AFM tip is similar (or smaller) in size than the tip itself. This is shown in Figs. S4b and S4c. This region, however, undergoes a sizeable broadening whenever the conductance at the surface increases due to the corresponding accumulation of charges (see Fig. S4d). This convolution between tip and domain-wall properties results in the bias-dependent broadening of the electrically dressed domain walls documented in Fig. 1 of the main text.

The aforementioned solution can be straightforwardly generalized to account for the redistribution of charges probed by the c-AFM tip when the *yz*-plane is scanned. We assume that the sample occupies the $y < 0$ region. Then, the landscape of conductivity associated with the domain wall can be taken as

$$\sigma = \begin{cases} \sigma_{DW} & -d < z < d \\ \sigma_{bulk} & otherwise \end{cases} \quad (S3)$$

where $\sigma_{DW}$ represents the nominal conductance of the domain wall, as described in the main text, while the effective width $w = 2d$ is defined such that:

$$d = \frac{\int_0^\infty dx\left(1-\exp(-\frac{\phi(x)}{k_BT})\right)}{1-\exp(-\frac{\phi_{DW}}{k_BT})} = \frac{\frac{\varepsilon}{ep_0}\int_0^\infty dx\phi''(x)}{1-\exp(-\frac{\phi_{DW}}{k_BT})} = \frac{\sqrt{\frac{\varepsilon k_BT}{e^2 p_0}}\Pi\cos\alpha}{1-\exp(-\frac{\phi_{DW}}{k_BT})} \quad (S4)$$

In this way, the hole density in the region $-d < z < d$ corresponds to the average density of holes depleted/accumulated by the bound charge at the domain wall. Thus, distribution of current injected by the c-AFM tip can be obtained from the potential

$$\phi_0(x,y,z) = \begin{cases} \dfrac{I}{2\pi\sigma_{bulk}} \sum_{n=0}^{-\infty} \dfrac{gf^n}{\sqrt{\rho^2+(z-2nd)^2}} & d < z \\ \dfrac{I}{2\pi\sigma_{DW}} \sum_{n=-\infty}^{\infty} \dfrac{f^{|n|}}{\sqrt{\rho^2+(z-2nd)^2}} & -d < z < d \\ \dfrac{I}{2\pi\sigma_{bulk}} \sum_{n=0}^{\infty} \dfrac{gf^n}{\sqrt{\rho^2+(z-2nd)^2}} & z < -d \end{cases} \quad (S5)$$

For a tip placed on top of the domain wall (say at $x = y = z = 0$) whose effective radius $r$ is much smaller than $d$, the conductance turns out to be

$$\sigma_{spread} \sim \dfrac{\sigma_{DW}}{1+\dfrac{r}{d}\ln\dfrac{\sigma_{DW}+\sigma_{bulk}}{2\sigma_{bulk}}}. \quad (S6)$$

This expression is obtained according to Eq. (S5) from the ratio between the total current injected by the tip, $I$, and the electric potential at the tip:

$$V \approx \dfrac{I}{2\pi\sigma_{DW}}\left(\dfrac{1}{r}+\dfrac{1}{d}\sum_{n=1}^{\infty}\dfrac{f^n}{n}\right) = \dfrac{I}{2\pi\sigma_{DW}r}\left(1+\dfrac{r}{d}\ln\dfrac{1}{1-f}\right) \quad (S7)$$

In this way, we account for the nontrivial spreading of current out of the domain walls as illustrated in Figs. 3c and 3d of the main text. Note that this spreading also affects the angular dependence of the domain wall conductance as indicated by the corresponding fit shown in Fig. 3a of the main text.